\documentclass[preprintnumbers,amsmath,amssymb]{revtex4}
\usepackage{graphics}
\usepackage{dcolumn}

\begin{document}
\title{Magnetic susceptibility of the square lattice Ising model}
\author{Tuncer Kaya}
\address{ Department of Physics, Yildiz Technical University,
        34220 Davutpa\c sa, Istanbul, Turkey}

\begin{abstract}
In this work, the susceptibility of the square lattice Ising model is investigated using the
recently obtained average magnetization
interrelation, which is given by $\langle\sigma_{0, i}\rangle=
\langle\tanh[K(\sigma_{1,i}+\sigma_{2,i}+\dots +\sigma_{z,i})+H]\rangle $. Here, $z$
is the number of nearest neighbors, $\sigma_{0,i}$ denotes the
central spin at the $i^{th}$ site while $\sigma_{l,i}$,
$l=1,2,\dots,z$, are the nearest neighbor spins around the central
spin, $K=J/(k_{B}T)$, where $J$ is the nearest neighbor coupling constant, $k_{B}$ is the Boltzmann's constant
and $T$ is the temperature of the system. In our investigation, inevitably we have to make a conjecture about
the three-site correlation function appearing in the obtained relation of this paper. The conjectured form of
the the three spin correlation function is given by the relation, $\langle\sigma_{1}\sigma_{2}\sigma_{3}\rangle=a(K,H)
\langle\sigma\rangle+[1-a(K,H)]\langle\sigma\rangle^{(1+\beta^{-1})}$. Here $\beta$ denotes the critical exponent for
the average magnetization and $a(K,H)$ is a function whose behavior will be described around the critical point with
an arbitrary constant. To elucidate the relevance of the method used in this paper, we have first calculated the
susceptibility of the 1D chain as an example, and the obtained susceptibility expression is seen as equivalent to
the result of the susceptibility of the conventional method. The magnetic critical exponent $\gamma$ of the square
lattice Ising model is obtained as $\gamma=1.72$ for $T\!>\!T_{c}$, and $\gamma=0.91$ for $T\!<\!T_{c}$.
\end{abstract}
\keywords{Ising model \sep Phase transition \sep Magnetic
susceptibility\sep Critical exponent}

\maketitle

\section{Introduction}
The square lattice Ising model is probably the most important statistical mechanical
model ever proposed in terms of its applications.

Since Onsager's \cite{Onsager} celebrated the solution of the Ising model free energy in 1944, followed by
\cite{Yang} the proof of Onsager's result for the spontaneous magnetization in 1952, physicists have devoted themselves
to the study of the problem of elucidating the susceptibility of the square lattice Ising model. As is well known, the
Onsager solution in 1944 was publicly presented without a proof \cite{ Guttmann}. Although there is no known
closed-form expression for the susceptibility $\chi$, a large body of knowledge about the susceptibility is available.
Attempting to list all these contributions is impossible in this paper.

We will, however only mention some of the research to indicate the main mathematical approaches in the investigation of
the square lattice Ising model. We refer the reader to McCoy's book \cite{McCoy} and the review article \cite{Coy} for
a review of these
developments. We crave the forgiveness of the researchers whose papers are not cited in this paper.

In 1976 Wu, McCoy, Tracy, and Barouch \cite{Wu} showed that the susceptibility could be expressed as an
infinite sum of $n$-dimensional integrals. The main result of their analysis can be given as the statement: for the
low-temperature susceptibility $(T<T_{c})$, only even powers of $n$ contribute, starting at $n = 2$, while for the
high-temperature susceptibility $(T>T_{c})$, only odd powers contribute.
In 1999, Nickel \cite{Nickel1,Nickel2} showed that the susceptibility has a natural
boundary in the complex plane and that such functions can not be $D$-finite or holonomic functions. Later it was proved
that each of these integrals $ \chi^{(n)}$ is $D$-finite but argued that their infinite sum, that is, the full
susceptibility $\chi$ itself, is not \cite{Orrick,Orrick2}. In a series of papers, Maillard and co-workers
\cite{Zenine1,Zenine2,Boukraa1,Nickel3,Bostan,Boukraa2} found the linear
ordinary differential equations (ODEs) satisfied by $\chi^{(3)},\dots, \chi^{(6)}$. In 2011, Chan and his
coworkers \cite{Chang} extended the work in Ref. \cite{Nickel1} to other two-dimensional lattices, and gave
an expansion of the scaled form of the susceptibility to unprecedented accuracy.

Almost half a century has passed since the formal expression for the magnetic susceptibility of square lattice Ising
model \cite{Coy} is obtained. There is still, however, so much discussion and uncertainty in the investigation of the
magnetic susceptibility of the square lattice Ising model.
Therefore, it is natural to ask the question: why it is so hard to obtain a closed expression for the susceptibility
in this model. To answer this question, it is proper to recall the basic definition of magnetic susceptibility, which
is given by the relation $\chi(T)=\frac{d\langle\sigma\rangle}{dh}|_{h=0}$. Here $\langle\sigma\rangle$ is the average
magnetization, $T$ denotes temperature, and $h$ is the external field. Since the average magnetization is known only in
zero magnetic fields (h=0), it is impossible to take the derivative appearing in the definition of magnetic
susceptibility.

Therefore, the susceptibility is usually studied through its relation with the zero-field spin-spin correlation function:
\begin{eqnarray*}
&&\chi(T)=\frac{1}{k_{B}NT}\sum_{i,j,k,l}\{\langle\sigma_{k,l}\sigma_{i,j}\rangle-\langle\sigma\rangle^{2}\}.
\end{eqnarray*}
Here, $i, j$ and $k, l $ denote lattice positions and run all over the system, $\langle\sigma\rangle$ is the
spontaneous magnetization. The correlation terms appearing the above general relation are approximated by replacing
the correlation function with $\langle\sigma_{0,0}\sigma_{i,j}\rangle$ or even with $\langle\sigma_{k,l}
\sigma_{i,i}\rangle$. $T_{c}$ denotes the critical temperature, and we recall that for the isotropic square
lattice Ising model, horizontal and vertical interaction constants have
the same value $J$. Then, the spontaneous magnetization is given for $T\! <\! T_{c}$ by
$\langle\sigma\rangle=(1-k_{1}^2)^{1/8}$.
Here $k_{1}=\textnormal{sinh}(2J/k_{B}T)^{-2}$ and $\langle\sigma\rangle$ is zero for $T\!>\!T_{c}$.

The correlation function appearing in the susceptibility relation
has been studied by many authors \cite{Kauf,Coy1,Coy2} and by
various methods. However, a compactly closed expression of any
$\langle\sigma_{0,0}\sigma_{i,j}\rangle$ have not been yet found
\cite{Yamada}, which indicates that the calculation of the square
lattice susceptibility is still an open problem.

Of course, applying daunting
mathematical treatments, the above-cited studies have made substantial advances in elucidating the properties of
the susceptibility of the square lattice Ising model. On the other hand, the mathematics used in these research
papers gets so complicated that it is impossible to follow up on the physical reasoning and even the mathematical
approximations used.

Therefore, it might be relevant and important to investigate the susceptibility of the square lattice Ising model
from a different perspective. To this end, we are going to exploit the recently obtained interrelation \cite{Kaya1}
for the average magnetization, which is given by the following expression as
\begin{eqnarray}
\langle\sigma\rangle=\langle\tanh[K(\sigma_{1}\cdots \sigma_{z})+H]\rangle.
\end{eqnarray}
Here, $\sigma_{1},\cdots,\sigma_{z}$ are the nearest neighbors spins around its central spin. By applying this
relationship together with a physically plausible assumption for odd spins correlation functions, we have obtained
almost exact average magnetization expressions for the 2D lattices \cite{Kaya2} and 3D cubic lattice \cite{Kaya3}
in the absence of external magnetic field.

We think it is worthwhile to mention that, in our recent papers, we have given the details of the calculations
(with some useful tricks and tools) allowing one to obtain average magnetization expressions for other 3D Ising
lattices. The obtain average magnetization expressions are almost equivalent to the exactly obtained results and
simulation data. In addition, in the calculation of the average magnetization relations, applied mathematics is
quite tractable and manageable.

Our primary goal in this paper is to calculate the susceptibility of the square lattice Ising model with the same
tractable and manageable mathematical procedures. In doing so, we would also like to gain some more insight into the
additional relevance of the conjectured assumption in our previous work. Here, the susceptibility of the square lattice
is going to be investigated by the same average magnetization interrelation and we will use the same mathematical form
of odd spins correlation function used in our previous studies. Of course, the odd spins correlation function is going
to be modified slightly to take into account the external magnetic field dependence.

To elucidate the mathematical procedure, we first calculate the susceptibility of the 1D Ising model as an example,
so that we can compare our result with the well-known exact susceptibility expression of the 1D Ising chain. In doing so,
we hope to provide adequate confirmation for the relevance of the method which will be used in the investigation of the
susceptibility of the square lattice Ising model.

This paper is organized as follows. In the next section, we are going to calculate the susceptibility of the 1D Ising
chain to elucidate and test the validity of the mathematical procedure which is going to be used in the calculation of
the susceptibility of the square Ising lattice. In the last section, the susceptibility of the square Ising lattice is
going to be calculated with the method developed for the treatment of the 1D chain. In doing so, we are going to try to
present our calculations and the necessary assumption as clearly as possible. In the same section, we are going to
discuss both the used method and some useful indications of possible future research problems which can be treated with
the method of this paper.

\section{The susceptibility of 1D Ising chain}
We start this section with the previously derived formula in reference \cite{Kaya1}, which is given by Eq. (1) in
the introduction.
To find the critical coupling strength, we need to consider $H=0$ case. So, taking into account Eq. (1) for 1D Ising
chain in the absence of external field and if the hyperbolic tangent function is expressed as
\begin{equation}
\tanh[K(\sigma_{1}+\sigma_{2})]=\frac{1}{2}(\sigma_{1}+\sigma_{2})\tanh(2K).
\end{equation}
The critical coupling strength can be readily obtained as follows. Substituting this relation into Eq. (1) leads to
$\langle \sigma\rangle= \langle\sigma\rangle \tanh(2K)$. From this equation, it is obvious that the non-zero values
for average magnetization $\langle\sigma\rangle $ is only possible at zero temperature. Here $K$ is defined as
$K=J/(k_B T)$, where $T$ is the temperature.
Thus, the critical temperature $T_{c}$ for 1D chain is equal to zero. Since the susceptibility is defined as the
derivative of the average magnetization with respect to the external magnetic field $h$, which is defined as
$H=h/(k_B T)$, we need to obtain the average magnetization of the 1D chain as a function of $H$. To this end,
we write the hyperbolic tangent function appearing in Eq. (1) as
\begin{equation}
\tanh[K(\sigma_{1}+\sigma_{2})+H]=C_{1}(\sigma_{1}+\sigma_{2})+C_{2}\sigma_{1}\sigma_{2}+C_{3}.
\end{equation}
Under the consideration of different orientations of $\sigma_{1}$ and $\sigma_{2}$, the coefficients $C_{1}$,
$C_{2}$ and $C_{3}$ can be obtained as,
\begin{eqnarray*}
&&\!\!\!\!\!\!\!\!\textstyle C_{1}(K,H)\!=\!\frac{1}{4}[\tanh(2K+H)]-\tanh(-2K+H), {}
\nonumber \\ &&
\!\!\!\!\!\!\!\!\textstyle C_{2}(K,H)\!=\!\frac{1}{4}[\tanh(2K+H)+\tanh(-2K+H)+2\tanh(H)]-\tanh(H),
{}\nonumber \\ &&
\!\!\!\!\!\!\!\!\textstyle C_{3}(K,H)\!=\!\frac{1}{4}[\tanh(2K+H)+\tanh(-2K+H)+2\tanh(H)].
\end{eqnarray*}
Substituting these final relations into Eq. (3), and then taking the average of the both side of the equation,
leads to
\begin{equation}
\langle\sigma\rangle=2C_{1}\langle\sigma\rangle+C_{2}\langle\sigma_{1}\sigma_{2}\rangle+C_{3}.
\end{equation}
Thus $\langle\sigma\rangle$ can be expressed as
\begin{equation}
\langle\sigma\rangle=\frac{C_{2}\langle\sigma_{1}\sigma_{2}\rangle+C_{3}}{1-2C_{1}}.
\end{equation}
The susceptibility is defined as
\begin{equation}
\chi(K,0)=\frac{d\langle\sigma\rangle}{dh}\Big|_{h=0} = \frac{1}{k_{B}T}\frac{d\langle\sigma\rangle}{dH}\Big|_{H=0}.
\end{equation}
After doing some algebra, $\chi(K,0)$ can be expressed as
\begin{equation}
\chi(K,0)=\frac{1}{k_{B}T}\frac{1+[\frac{1}{2}+\langle\sigma_{1}\sigma_{2}\rangle]
\textnormal{sech}^{2}(2K)}{1-\tanh(2K)}.
\end{equation}
For large values of $K$ or small values of $T$, this equation can be expressed as
\begin{equation}
\chi(K,0)=\frac{1}{k_{B}T}e^{4J/k_B T}.
\end{equation}
This final susceptibility relation is equivalent to the susceptibility relation obtained in \cite{Pathria}.
At this point it is important to point out that we have obtained the susceptibility relation without using
the well known 1D chain average magnetization relation, which is given by the following relation as
\begin{equation}
\langle\sigma\rangle=\frac{\textnormal{sinh}(H)}{[e^{-4K}+\textnormal{sinh}^{2}(H)]^{\frac{1}{2}}}.
\end{equation}

In the next section, our primary goal will be to calculate the susceptibility of the square lattice Ising model
in the same manner used in this section.
Exploiting these advantages of mathematical simplicity and tractability, we will make use of the available average
magnetization relation for the square lattice. We will work with almost the same mathematical form for the three
spins correlation function conjectured previously in reference \cite{Kaya2}. Focusing on the transition region,
we can calculate the susceptibility relation for the square lattice and we are also able to extract the values of
the magnetic critical exponent $\gamma$ for $K\!<\!K_{c}$ and $K\!>\!K_{c}$ for the square lattice.

In addition, we also give reasonable discussion for the modification of the conjectured form of the three spins
correlation function in the presence of an external magnetic field. To better understand this connection between
the three spins correlation function in the presence and the absence of an external magnetic field, we trust the
relevance of the previously obtained average magnetization relations obtained in references \cite{Kaya2,Kaya3}
in which the same mathematical form of the three spin correlation function was used.

\section{The susceptibility of the square lattice Ising Lattice}
In this section we apply the same method used in the previous section to calculate the susceptibility of the
square Ising lattice. As it is done above, the procedure starts by writing the Eq. (1) for square lattice as
\begin{equation}
\langle \sigma_{0,i}\rangle=
\langle \tanh[K(\sigma_{1,i}+\sigma_{2,i}+\sigma_{3,i} +\sigma_{4,i})+H]\rangle,
\end{equation}
where $K$ is the coupling strength and $\sigma_{0,i}$ denotes the
central spin at the $i^{th}$ site, while $\sigma_{l,i}$,
$l=1,2,3,4$, are the nearest neighbor spins around the central
spin. In what follows the index $i$ is not used since it is not necessary in the treatment. We express the
hyperbolic tangent function with the following equivalent relation as
\begin{eqnarray}
&&\tanh[K(\sigma_{1}+\sigma_{2}+\sigma_{3}+\sigma_{4})+H]=A_{1}(\sigma_{1}+\sigma_{2}+\sigma_{3}+\sigma_{4}){}
\nonumber \\ && +A_{2}(\sigma_{1}\sigma_{2}+\sigma_{2}\sigma_{3}+\sigma_{3}\sigma_{4}+\sigma_{4}\sigma_{1}+
\sigma_{1}\sigma_{3}+\sigma_{2}\sigma_{4}){}\nonumber \\ &&+A_{3}(\sigma_{1}\sigma_{2}\sigma_{3}+\sigma_{2}
\sigma_{3}\sigma_{4}+\sigma_{3}\sigma_{4}\sigma_{1}+\sigma_{4}\sigma_{1}\sigma_{2})+A_{4}
\sigma_{1}\sigma_{2}\sigma_{3}\sigma_{4}+A_{5},
\end{eqnarray}
\begin{eqnarray*}
&&\!\!\!\!\!\!\!\!\textstyle A_{1}\!=\!\frac{1}{16}[\tanh(4K\!+\!H)\!-\!\tanh(-4K\!+\!H)\!+\!2\tanh(2K\!+\!H)\!-
\!2\tanh(2K\!\!+\!\!H)],{}
\nonumber \\ &&
\!\!\!\!\!\!\!\!\textstyle A_{2}\!=\!\frac{1}{16}[\tanh(4K\!+\!H)\!+\!\tanh(-4K\!+\!H)\!-\!2\tanh(H)],{}\nonumber \\ &&
\!\!\!\!\!\!\!\!\textstyle A_{3}\!=\!\frac{1}{16}[\tanh(4K\!+\!H)\!-\!\tanh(-4K\!+\!H)\!-\!2\tanh(2K\!+\!H)\!+
\!2\tanh(2K\!\!+\!\!H)],{}
\nonumber \\ &&
\!\!\!\!\!\!\!\!\textstyle A_{4}=2A_{2}-A_{5}+\tanh(H),{}\nonumber \\ &&
\!\!\!\!\!\!\!\!\textstyle A_{5}=A_{2}+\frac{1}{4}[\tanh(2K+H)+\tanh(-2K+H)+2\tanh(H)],
\end{eqnarray*}
\begin{eqnarray}
\langle \sigma \rangle=4A_{1} \langle \sigma \rangle\!+\!A_{2}[4\langle \sigma_{1}\sigma_{2} \rangle\!+\!2\langle
\sigma_{1}\sigma_{3} \rangle]\!+\!4A_{3}\langle \sigma_{1}\sigma_{2}\sigma_{3} \rangle\!+\!A_{4}\langle \sigma_{1}
\sigma_{2}\sigma_{3}\sigma_{4} \rangle\!\!+\!\!A_{5}.
\end{eqnarray}
The three spin correlation function was conjectured as in references \cite{Kaya2,Kaya3} as
\begin{eqnarray}
\langle \sigma_{1}\sigma_{2}\sigma_{3} \rangle=a(K_{c},0)\langle \sigma \rangle+[1-a(K_{c},0)]\langle \sigma
\rangle^{\frac{1+\beta}{\beta}},
\end{eqnarray}
in the absence of external magnetic field. It is important to notice that the value of $a(K{c},0)$ was obtained
in our previous paper \cite{Kaya2}.

In this paper, we assume that the form of the three spin correlation function in the absence of external field
can also be the relevant relation for the three spin correlation function in the presence of external magnetic field. This can be expressed as,
\begin{eqnarray}
\langle \sigma_{1}\sigma_{2}\sigma_{3} \rangle=a(K,H)\langle \sigma \rangle+[1-a(K,H)]\langle \sigma
\rangle^{\frac{1+\beta}{\beta}}.
\end{eqnarray}
Thus, substituting this relation into Eq. (12) leads to
\begin{eqnarray}
f_{1}(K,H)\langle\sigma\rangle+f_{2}(K,H)\langle \sigma \rangle^{\frac{1+\beta}{\beta}}=f_{3}(K,H).
\end{eqnarray}
The functions appearing in this equation are given by
\begin{eqnarray*}
&&\!\!\!\!\!\!\!\!\textstyle f_{1}(K,H)\!=\!1-4A_{1}(K,H) -4A_{3}(K,H)a(K,H),{}
\nonumber \\ &&
\!\!\!\!\!\!\!\!\textstyle f_{2}(K,H)\!=\!-4A_{3}[1-a(K,H)],{}\nonumber \\ &&
\!\!\!\!\!\!\!\!\textstyle f_{3}(K,H)\!=\!A_{2}[4\langle \sigma_{1}\sigma_{2} \rangle+
2\langle \sigma_{1}\sigma_{3} \rangle]+A_{4}(K,H)\langle \sigma_{1}\sigma_{2}\sigma_{3}\sigma_{4} \rangle+A_{5}(K,H).
\end{eqnarray*}

Taking the derivative of the both sides of the Eq. (15) with respect to the external field $h$, the following
expression for the susceptibility of square lattice obtained as,
\begin{eqnarray}
\lim_{H\to 0} \chi(K,H)=\frac{1}{k_{B}T} \lim_{H \to 0}\frac{I_{1}(K,H)}{I_{2}(K,H)}.
\end{eqnarray}
Here $I_{1}$ and $I_{2}$ are given by the relations,
\begin{eqnarray*}
&&I_{1}(K,H)=\frac{df_{3}}{dH}-\langle \sigma \rangle \frac{df_{1}}{dH}-\langle \sigma \rangle^{\frac{1+\beta}{\beta}}
\frac{df_{2}}{dH},{}\nonumber \\&&
I_{2}(K,H)=f_{1}+\frac{1+\beta}{\beta}\langle \sigma \rangle^{\frac{1}{\beta}}f_{2},
\end{eqnarray*}
\begin{eqnarray*}
\lim_{H \to 0} \frac{df_{1}}{dH}=-\frac{1}{4}[2\tanh(4K)-4\tanh(2K)]\frac{a(K,H)}{dH}\!\!\mid_{H=0},
\end{eqnarray*}
\begin{eqnarray*}
\lim_{H \to 0} \frac{df_{2}}{dH}=\frac{1}{4}[2\tanh(4K)-4\tanh(2K)]\frac{a(K,H)}{dH}\!\!\mid_{H=0},
\end{eqnarray*}
\begin{eqnarray*}
\lim_{H \to 0} \frac{df_{3}}{dH}&=&\frac{1}{8}\tanh(4K)^{2}[4\langle \sigma_{1}\sigma_{2} \rangle \!\!\mid_{H=0}+
2\langle \sigma_{1}\sigma_{3}\rangle\!\!\mid_{H=0}] {}\nonumber\\&&+\frac{1}{8}[3-4\textnormal{ sech}(2K)^2+
\textnormal{sech}(4K)^2 ] \langle \sigma_{1}\sigma_{2}\sigma_{3}\sigma_{4} \rangle\!\!\mid_{H=0},
\end{eqnarray*}
\begin{eqnarray*}
\textstyle f_{1}(K,0)\!\!&=&1-\frac{1}{4}[4\tanh(2K)\!+\!2\tanh(4K)]\!{}\nonumber \\&&-\frac{1}{4}[2\tanh(4K)-
4\tanh(2K)]a(K,0),
\end{eqnarray*}
\begin{eqnarray*}
\textstyle f_{2}(K,0)=-\frac{1}{4}[-4\tanh(2K)+2\tanh(4K)][1-a(K,0)].
\end{eqnarray*}
For $K<K_{c}$, using the above relations, the susceptibility can be expressed as
\begin{eqnarray*}
\chi(K,0)=\frac{\frac{1}{8}\tanh(4 K)^{2}C+\frac{1}{8}[3-4 \textnormal{sech}(2K)^2+\textnormal{sech}(4K)^2 ]
\langle \sigma_{1}\sigma_{2}\sigma_{3}\sigma_{4} \rangle\!\!\mid_{H=0} }{1-[1-a(K,0)]\tanh(2K)-\frac{1}{2}[1+
a(K,0)]\tanh(4K)}.
\end{eqnarray*}
Here $C$ is defined as $C=[4\langle \sigma_{1}\sigma_{2} \rangle\!\!\mid_{H=0}+2\langle \sigma_{1}\sigma_{3}
\rangle\!\!\mid_{H=0}]$.
Now, if we can figure out the form of the unknown function $a(K,0)$, then we can calculate the susceptibility
for $K\!<\!K_{c}$.
\begin{figure*}[!hbt]
\begin{center}
\scalebox{1.1}{\includegraphics{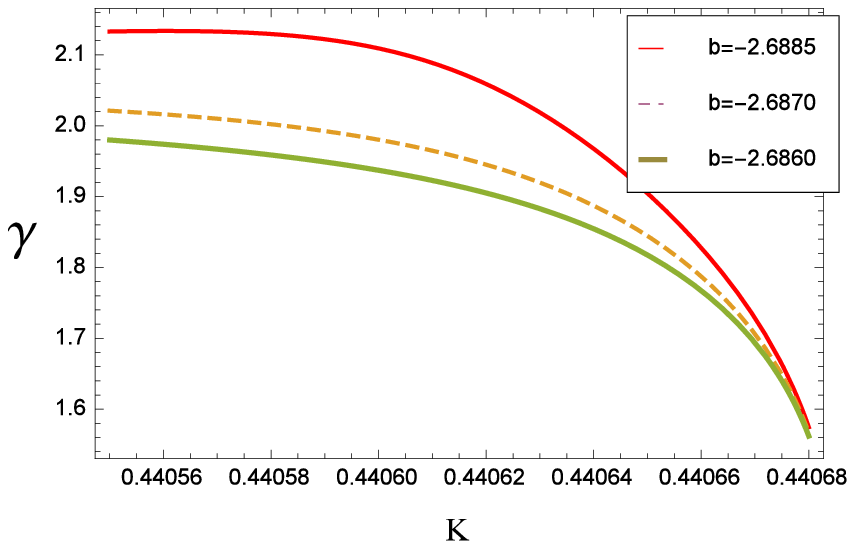}}
\end{center}
\caption{The values of $\gamma$ for the case $K<K_{c}$ in the
vicinity of $K_{c}$ for the three different values of the
arbitrary parameter $b$, which is chosen as $ b\simeq -2.68$.}
\end{figure*}
\begin{figure*}[!hbt]
\begin{center}
\scalebox{1.1}{\includegraphics{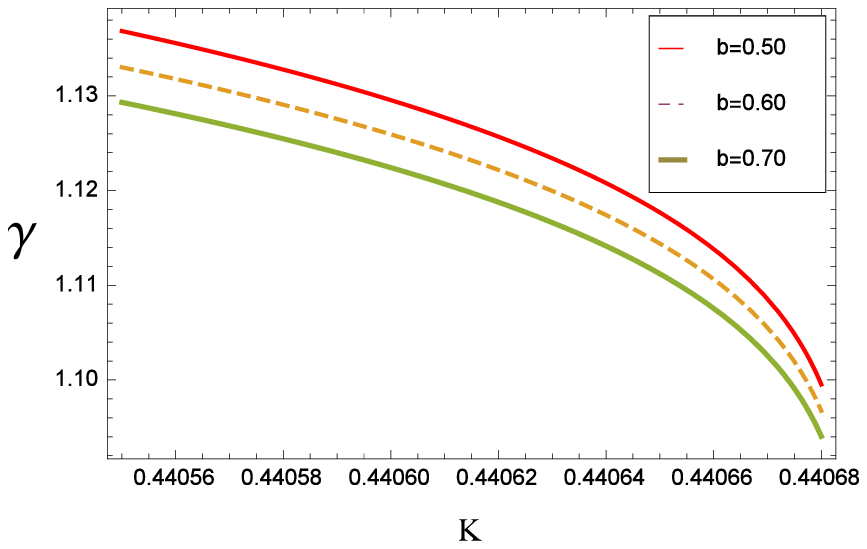}}
\end{center}
\caption{The values of $\gamma$ for the case $K<K_{c}$ in the
vicinity of $K_{c}$ for the three different values of the
arbitrary parameter $b$. }
\end{figure*}
\begin{figure*}[!hbt]
\begin{center}
\scalebox{1.1}{\includegraphics{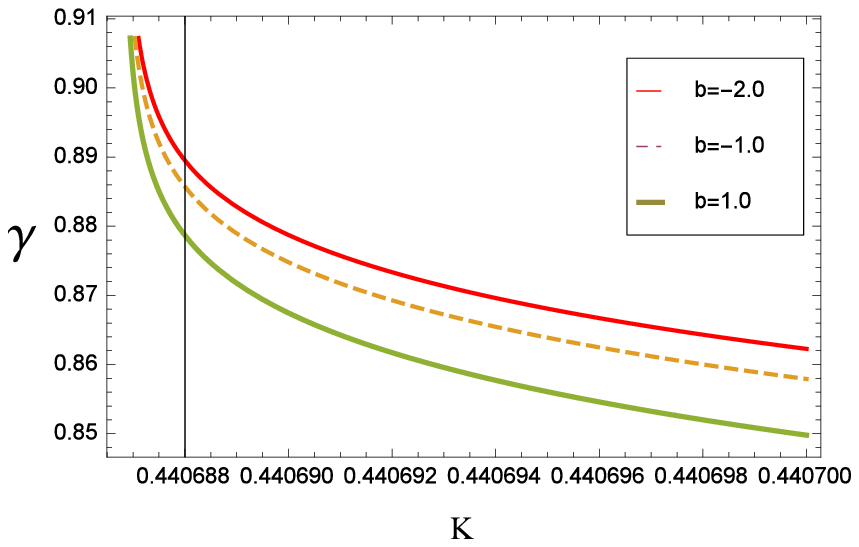}}
\end{center}
\caption{The values of $\gamma$ for the case $K>K_{c}$ in the
vicinity of $K_{c}$ for the three different values of the
arbitrary parameter $b$. }
\end{figure*}
\begin{figure*}[!hbt]
\begin{center}
\scalebox{1.1}{\includegraphics{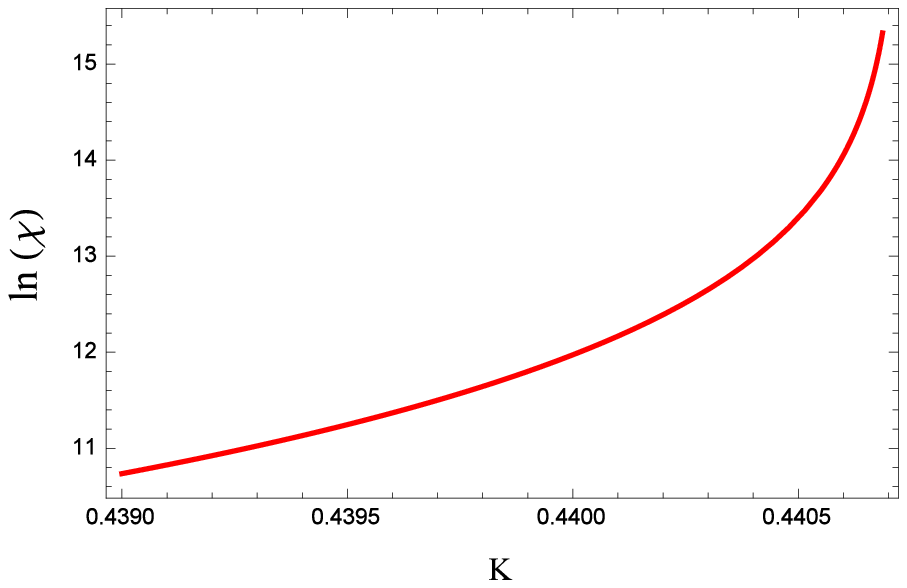}}
\end{center}
\caption{The plot of $\ln(\chi)$ for $K<K_{c}$ in the vicinity of
$K_{c}$. }
\end{figure*}
\begin{figure*}[!hbt]
\begin{center}
\scalebox{1.1}{\includegraphics{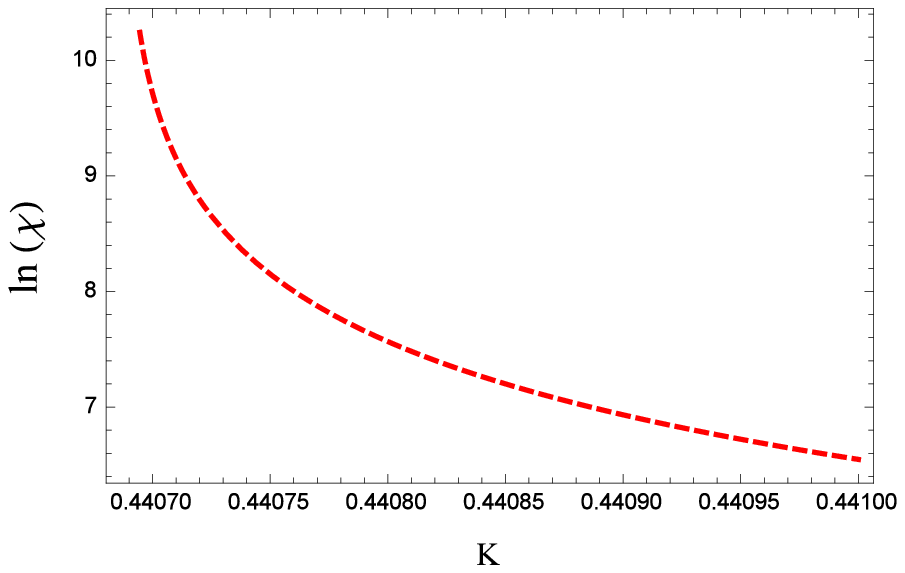}}
\end{center}
\caption{The plot of $\ln(\chi)$ for $K>K_{c}$ in the vicinity of
$K_{c}$. }
\end{figure*}

To this end we are going to exploit $a(K_{c},0)=0.75736$ which was
obtained in reference \cite{Kaya2}. Since our purpose is to obtain
the susceptibility function for values near to $K_{c}$, we can
expand $a(K,0)$ around $K_{c}$ as
\begin{eqnarray}
a(K,0)=a(K_{c},0)+b(K-K_{c})+\cdots,
\end{eqnarray}
here $b$ is equal to $\frac{da(K,0)}{dK}|_{K=K_{c}}$. After defining $a(K,0$), we are ready to investigate
the expression for susceptibility. The first thing one may notice is that all of the terms in the numerator
of $\chi(K,0)$ are smooth functions and they all have finite values for the any values of $K$ while the
denominators go to zero (or they assume very small values) around the critical point.

This means that the behavior of $\chi$ strongly depends on the values of denominator and it is independent of
the finite values of numerator. If the
scaling form of susceptibility, which is defined as $\chi=(1-\frac{K}{K_{c}})^{-\gamma}$, is recalled, the
critical exponent $\gamma$ can be given by the following relation as
\begin{eqnarray}
\gamma=\frac{\ln[1-[1-a(K,0)]\tanh(2K)-\frac{1}{2}[1+a(K,0)]\tanh(4K)]}{\ln(K_{c}-K)}.
\end{eqnarray}
In Fig. 1 and Fig. 2, the critical exponent $\gamma$ is plotted for different values of the parameter $b$
introduced to describe $a(K,0)$. From these figures, it is obvious that the values of $\gamma$ strongly depend
on the values of $b$ and the values of $(K-K_{c})$. It is clear from Fig. 1 that the values of $\gamma$ approach
a definite value when the values of $K$ are very close to $K_{c}$. In this limit, the value of $\gamma$ is
around $\gamma=1.7$ which is in agreement with the value $\gamma=1.75$ obtained from the scaling theory. Fig. 2 is
plotted for positive values of the parameter $b$. As seen from the figure, the values of $\gamma$ are
around $\gamma=1.1$ when $K$ is very close to $K_{c}$. Now, we are going to investigate the behavior of
the critical exponent $\gamma$ for the values of $K$ greater than $K_{c}$ in the same manner as it is treated
in the $K\!<\!K_{c}$ case. The first thing one can notice is that the terms in $I_{1}$ are smooth and they all
have finite values for all the values of $K$. This property of $I_{1}$ easily leads to the following $\gamma$ relation,
\begin{eqnarray}
\gamma=\frac{\ln[f_{1}(K,0)+9[1-\textnormal{sinh}(2K)^{-4}]f_{2}(K,0)]}{\ln(K_{c}-K)},
\end{eqnarray}
where, for the average magnetization we used the Onsager relation, $\langle\sigma\rangle=
[1-\textnormal{sinh}(2K)^{-4}]^{1/8}$, and the critical exponent $\beta$ of square lattice is equal to $1/8$.

The obtained $\gamma$ relation for the case $K\!\!>\!\!K_{c}$ is
plotted in Fig. 3. From this figure, it is obvious that the values
of $\gamma$ around the critical point $K_{c}$ do not depend on the
values of the arbitrary parameter $b$. And the values of $\gamma$
approach $0.91$ regardless of the values of the arbitrary
parameter $b$.

To elucidate the general feature of the behavior of magnetic susceptibility, the logarithm of the magnetic
susceptibility relations obtained for the cases $K\!\!<\!\!K_{c}$ and $K\!\!>\!\!K_{c}$ in this paper for the
square lattice Ising model are plotted in Fig. 4 and Fig. 5 respectively.

\section{Conclusion and Discussion}
In this paper, a new approach is introduced to investigate and calculate the susceptibility of the square
lattice Ising model. To do so, we start our investigation by applying the exact interrelation for the average
magnetization introduced previously. The application of this
interrelation in the presence of an external magnetic field produces the relations which relate average magnetization
of the lattices to the even spins and the odd spins correlation functions.

We have assumed that the even spins correlation functions behave smoothly and
continuously around the critical point. As a result, the derivative of the correlation functions with respect to
the external magnetic field gives finite values at the critical point for $H=0$. We have also assumed that the
three spin correlation function obeys the same mathematical form as conjectured in our previous works by just
replacing the constant $a$ with a function $a(K, H)$. Exploiting the previously obtained result $a(K,0)$ at
critical point, $a(K,0)$ is expanded into series around $K=K_{c}$. Since we aim to calculate the susceptibility
of the square lattice for values of $K$ near $K_{c}$, the function $a(K,0)$ can be approximated by keeping just
two terms in the expansion.

We have observed that the values of the critical exponent $\gamma$ are strongly dependent on the values of the
arbitrary parameter for the case $K<K_{c}$, but the values of $\gamma$ converge to a fixed value if the value of
arbitrary parameter takes a value around $-2.65$. Using this consideration, we have determined the values
of $\gamma=1.72$ for $K<K_{c}$. On the other hand, for the values of $K$ is greater than $K_{c}$, the values
of $\gamma$ are slightly dependent on the values of the arbitrary parameter.

In the end, $\gamma$ converges to $0.91$. We believe that the approach used in this paper is quite relevant
and important. The method can be also applicable to the more complicated untouchable problem such as calculating
the susceptibility of the 3D cubic lattice Ising model. These types of problems are going to be the subject of
our future research.


\begin{thebibliography}{00}
\bibitem{Onsager} L. Onsager, Crystal statistics. I. A two-dimensional model with an order-disorder transition, Phys. Rev. {\bf 65} (1944) 117-149.
\bibitem{Yang}  C. N. Yang, The spontaneous magnetization of a two-dimensional Ising model, Phys. Rev. {\bf 85} (1952) 808-816.
\bibitem{Guttmann} A.J. Guttmann, I.Jensen, J.-M. Maillard, J. Pantone, 2016. Is the full susceptibility of the square-lattice Ising model a differentially algebraic function?, J. Phys. A Math. Theor. {\bf 49} 504002.
\bibitem{McCoy} B.M. McCoy, Advanced Statistical Mechanics, Oxford University Press, 2010.
\bibitem{Coy} B.M. McCoy, M. Assis, S. Boukraa, S. Hassani, J.-M. Maillard, W.P. Orrick, N. Zenine, The saga of the Ising susceptibility, in: B. Feigin, M. Jimbo, M. Okado (Eds.), New trends in quantum integrable systems, Proceedings of the Infinite Analysis 09, Kyoto, Japan, 2010, pp. 287–306.
\bibitem{Wu}T.T. Wu, B.M. McCoy, C.A. Tracy, E. Barouch, Spin-spin correlation functions for the
two-dimensional Ising model: Exact theory in the scaling region, Phys. Rev. B {\bf 13} (1976) 316-374.
\bibitem{Nickel1} B. Nickel, On the singularity structure of the 2D Ising model susceptibility, J. Phys. A: Math. Gen. {\bf 32} (1999) 3889-3906.
\bibitem{Nickel2} B. Nickel, Addendum to ''On the singularity structure of the 2D Ising model'', J. Phys. A: Math. Gen. {\bf 33} (2000) 1693-1711.
\bibitem{Orrick} W.P. Orrick, B. Nickel, A.J. Guttmann, J.H.H. Perk, The susceptibility of the square lattice Ising model: New developments,
 J. Stat. Phys. {\bf 102} (2001) 795-841.
\bibitem{Orrick2} W.P. Orrick, B.G. Nickel, A.J. Guttmann, J.H.H. Perk, Critical behavior of the two-dimensional Ising susceptibility,
Phys. Rev. Lett. {\bf 86} (2001) 4120-4123.
\bibitem{Zenine1} N. Zenine, S. Boukraa, S. Hassani, J.-M. Maillard, The Fuchsian differential equation of the
square lattice Ising model $\chi^{(3)}$ susceptibility, J. Phys. A: Math. Gen. {\bf 37} (2004) 9651-9668.
\bibitem{Zenine2} N. Zenine, S. Boukraa, S. Hassani, J.-M. Maillard, Square lattice Ising model susceptibility:
connection matrices and singular behaviour of $\chi^{(3)}$ and $\chi^{ (4)}$, J. Phys. A: Math. Gen. {\bf 38} (2005) 9439-9474.
\bibitem{Boukraa1} S. Boukraa, A.J. Guttmann, S. Hassani, I. Jensen, J.-M. Maillard, B. Nickel, N. Zenine, 2008.
Experimental mathematics on the magnetic susceptibility of the square lattice Ising model,
J. Phys. A: Math. Theor. {\bf 41} 455202.
\bibitem{Nickel3} B. Nickel, I. Jensen, S. Boukraa, A.J. Guttmann, S. Hassani, J.-M. Maillard, N. Zenine, 2010.
Square lattice Ising model $\tilde\chi ^{(5)}$ ODE in exact arithmetic, J. Phys. A: Math. Theor. {\bf 42} 195205.
\bibitem{Bostan} A. Bostan, S. Boukraa, A.J. Guttmann, S. Hassani, I. Jensen, J.-M. Maillard, N. Zenine, 2009.
High order Fuchsian equations for the square-lattice Ising model: $\tilde \chi ^{(5)}$, J. Phys. A: Math.
Theor. {\bf 42} 275209.
\bibitem{Boukraa2}S. Boukraa, S. Hassani, I. Jensen, J.-M. Maillard, N. Zenine, 2010. High order Fuchsian equations
for the square lattice Ising model: $\chi^{6}$,  J. Phys. A: Math. Theor. {\bf 43} 115201.
\bibitem{Chang} Y. Chan, A.J. Guttmann, B.G. Nickel, J.H.H. Perk, The Ising susceptibility scaling
function, J. Stat. Phys. {\bf 145} (2011) 549-590.
\bibitem{Kauf} B. Kaufman, L. Onsager, Crystal Statistics. III. Short-Range Order in a Binary Ising Lattice, Phys. Rev. {\bf 76} (1949) 1244-1252.
\bibitem{Coy1} B. McCoy, T.T. Wu, The Two Dimensional Ising Model, Harvard University Press, Cambridge, 1973.
\bibitem{Coy2} B. McCoy, 2010. Ising model: exact results, Scholarpedia, 5(7):10313, doi:10.4249/scholarpedia.10313.
\bibitem{Yamada}K. Yamada, Pair correlation function in the Ising square lattice, Prog. Theor. Phys. {\bf 76} (1986) 602-612.
\bibitem{Kaya1}T. Kaya, Exact three spin correlation function relations for the square and
the honeycomb Ising lattices, Chin. J. Phys. {\bf 66} (2020) 415-421.
\bibitem{Kaya2}T. Kaya, Relevant alternative analytic average magnetization calculation method for the square and the honeycomb Ising lattices, Chin. J. Phys. {\bf 77} (2022) 747-752.
\bibitem{Kaya3}T. Kaya, Relevant spontaneous magnetization relations for the triangular and the cubic lattice Ising model, Chin. J. Phys.
(2022), https://doi.org/10.1016/j.cjph.2022.03.043.
\bibitem{Pathria} R.K. Pathria, P.D. Beale, Statistical Mechanics, third ed., Elsevier, New York, 2011, pp. 478.
\end{thebibliography}
\end{document}